\documentclass[doublecol]{epl2} 
\usepackage{bm}
\usepackage{amsmath}
\usepackage{amssymb}

\title{Localization and diffusion of tracer particles in viscoelastic media with active force dipoles} 
\shorttitle{Localization and diffusion of tracer particles in viscoelastic media}

\author{Kento Yasuda,\inst{1} Ryuichi Okamoto,\inst{1} Shigeyuki Komura\inst{1}
and Alexander S. Mikhailov\inst{2,3}}
\shortauthor{K. Yasuda \textit{et al.}}

\institute{
\inst{1} Department of Chemistry, Graduate School of Science and Engineering,
Tokyo Metropolitan University, Tokyo 192-0397, Japan \\
\inst{2} Department of Physical Chemistry, Fritz Haber Institute of the Max Planck Society, 
Faradayweg 4-6, 14195 Berlin, Germany \\
\inst{3}Department of Mathematical and Life Sciences, Hiroshima University, Hiroshima 739-8526, Japan
}


\abstract{
Optical tracking \textit{in vivo} experiments reveal that diffusion of particles in biological cells is strongly enhanced in the presence of ATP and the experimental data for animal cells could previously be reproduced within a phenomenological model of a gel with myosin motors acting within it
[EPL \textbf{110}, 48005 (2015)]. 
Here, the two-fluid model of a gel is considered where active macromolecules, described as force dipoles, cyclically operate both in the elastic and the fluid components. 
Through coarse-graining, effective equations of motions for idealized tracer particles displaying local deformations and local fluid flows are derived. 
The equation for deformation tracers coincides with the earlier phenomenological model and thus confirms it. 
For flow tracers, diffusion enhancement caused by active force dipoles in the fluid component, and thus due to metabolic activity, is found. 
The latter effect may explain why ATP-dependent diffusion enhancement could also be observed in bacteria that lack molecular motors in their skeleton or when the activity of myosin motors was chemically inhibited in eukaryotic cells.
}
\begin{document}

\maketitle

\section{Introduction}

High-precision optical tracking experiments demonstrate that diffusion of microinjected or natural particles is strongly enhanced in the presence of ATP and, generally, under metabolic activity in the cytoplasm of a biological cell~\cite{Parry,Guo}. 
Similar behavior was previously found for diffusion within the nucleoplasm~\cite{Weber,Bruinsma14}. 
Thus, non-thermal active noises apparently dominate transport processes inside living cells. 
Two possible explanations have been proposed. 
The first of them relates the observed effects to activity of myosin motors that cross-link actin polymer filaments in eukaryotic (animal) cells~\cite{Levine,Levin1}. 
Importantly, strong diffusion enhancement in the presence of ATP could also be observed \textit{in vitro} in the experiments with 
synthetic actin-myosin systems~\cite{Misuno}. 
Another possibility is that non-thermal noise is of metabolic origin and it arises from fluid agitation caused by active conformational changes in protein molecules within the cytoplasm~\cite{Parry,Kapral}. 
Such explanation is supported by experimental results for bacteria (i.e., prokaryotic microorganisms) that lack molecular motors in their skeleton but where strong diffusion enhancement is nonetheless observed~\cite{Parry} 
(and ATP-dependent diffusion enhancement was also found when the skeleton had been dissolved~\cite{Parry}). 
Recently, it was moreover demonstrated that both the tracking and the microrheology data on active diffusion in eukaryotic cells can be reproduced in the framework of a phenomenological model for the motion of tracer particles trapped inside an elastic environment and subject only to active forces generated within it~\cite{Fodor}.

In this Letter, the problem is analyzed within a general model of an viscoelastic medium that combines strain fluctuations caused by molecular motors in the gel and mechanical agitation of the fluid component caused by metabolic activity within it. 
The classical two-fluid description is employed for the viscoelastic medium~\cite{DeGennes} and molecular motors, enzymes and other protein machines are modeled as active force dipoles. 
We introduce two kinds of hypothetical non-invasive tracers that are intended to visualize, respectively, elastic deformations and fluid flows. 
Reduced equations for random motion of both kinds of particles are derived. 
As we find, the derived equation for deformation tracers coincides with the previously proposed phenomenological model~\cite{Fodor}. 
On the other hand, relative random motion of flow tracers with respect to the elastic component is described by a Langevin equation with non-thermal metabolic noise.

Our analysis reveals that localization and diffusion phenomena are generally involved. 
The motion of tracers immobilized within the elastic subsystem is localized in the long-time limit, but it can show a diffusion-like behavior at the intermediate time scales shorter than the cooperative correlation times of molecular motor aggregates (i.e., myosin microfilaments) operating in the active gels of eukaryotes, as also found in the phenomenological model~\cite{Fodor}. 
On the other hand, motion of flow tracers remains diffusive in the long-time limit, and such active diffusion is controlled by metabolic processes that take place both in prokaryotes and eukaryotes. 
Assuming that the motion of real tracer particles represents repeated attachment and detachment to the polymer network, 
we expect for it a combination of localization and diffusion effects.

\section{The model}

Our study is based on the two-fluid model of gels~\cite{DeGennes}. 
A viscoelastic medium is described by the distributions of elastic deformations $\mathbf{u}(\mathbf{r} ,t)$ and fluid velocities $\mathbf{v}(\mathbf{r} ,t)$ within it. 
Neglecting inertial effects that do not play a role on the considered time scales, the evolution of the two fields is governed by 
\begin{equation}0 =\mu  \nabla ^{2}\mathbf{u} +\left (\mu  +\lambda \right ) \nabla \left ( \nabla  \cdot \mathbf{u}\right ) -\Gamma \left(\frac{ \partial \mathbf{u}}{ \partial t} -\mathbf{v}\right) +\mathbf{f}^{u}, 
\label{fluid2}
\end{equation}
\begin{equation}0 =\eta  \nabla ^{2}\mathbf{v} - \nabla p -\Gamma \left (\mathbf{v} -\frac{ \partial \mathbf{u}}{ \partial t}\right ) +\mathbf{f}^{v}. 
\label{fluid3}
\end{equation}
If the volume fraction of the elastic component is small, the conservation equation 
$\nabla  \cdot \mathbf{v} =0$ should additionally hold. 
Here, $\mu $ and $\lambda $ are the Lam\'{e} coefficients of the elastic network,  $\eta $ is the fluid viscosity, 
and $p$ is the pressure field that can be eliminated by using the above incompressibility condition. 
The coefficient $\Gamma$ specifies the relative strength of viscous friction forces when an elastic network is dragged by the fluid that flows through it. The approximate continuum description for a polymer gel holds only on the length scales that are much larger than the mesh size which is usually close  in its magnitude to the characteristic length 
$\ell =\left (\eta /\Gamma \right )^{1/2}$. 
Moreover, a characteristic viscoelastic relaxation time $\tau  =\eta /\mu$ can be also defined. 
Equations~(\ref{fluid2}) and (\ref{fluid3}) include forces $\mathbf{f}^{u}$ and $\mathbf{f}^{v}$ that are applied, respectively, to the elastic and the fluid components. 
Since we are only interested in non-thermal fluctuations, we do not include into such forces the thermal noise.

To study diffusion in viscoelastic biological systems, tracer particles with diameters $a$ significantly larger than the mesh size are employed ($a \gg \ell$). 
The dynamics of such particles inside a gel is itself a complicated problem; it depends on the mechanical properties of such particles and on their interactions with the polymer network~\cite{Diamant15}. 
Moreover, the presence of big particles will also modify the local dynamics of the gel because, for instance, the assumption of the small volume fraction of the non-fluid component will not be satisfied at the location of such a particle.

To avoid these complications, motions of \textit{non-invasive} flow velocity and deformation tracers will be considered by us.
A flow velocity tracer can be introduced by choosing a certain volume element of the fluid and following its motion with time. 
Similarly, a deformation tracer represents a chosen volume element of the elastic component.
The tracers can only follow those velocity and deformation fluctuations whose characteristic length scale is larger than the tracer size $a$. 
Hence, they visualize the coarse-grained fluctuating fields obtained by averaging over the volume elements of linear size $a$. 
Such fields include only the contributions from spatial Fourier modes with wavenumbers smaller than $q_{\rm c} =4\pi /a$.
The coarse-grained velocity and deformation fields are defined as
$V_{\alpha }(\mathbf{r} ,t) =(2\pi)^{-3} \int _{\left \vert \mathbf{q}\right \vert  <q_{\rm c}}
{\rm d} \mathbf{q} \, v_{\alpha }(\mathbf{q} ,t)e^{-i\mathbf{q}\cdot \mathbf{r}}$ and 
$U_{\alpha }\left (\mathbf{r} ,t\right ) =(2\pi)^{-3}\int _{\left \vert \mathbf{q}\right \vert  <q_{\rm c}}
{\rm d}\mathbf{q}\, u_{\alpha }\left (\mathbf{q} ,t\right )e^{-i\mathbf{q}\cdot \mathbf{r}}$,
respectively.

Generally, both compression and shear deformations will be induced.
We shall however only consider shear deformations because they can be visualized by incompressible tracers. 
Retaining only such deformations and performing a transformation to the spatial Fourier components of involved fields, we obtain from eqs.~(\ref{fluid2}) and (\ref{fluid3}) 
\begin{align}
0 = & -\mu q^{2}u_{\alpha }\left (\mathbf{q} ,t\right ) -\Gamma \left( \frac{\partial u_{\alpha }(\mathbf{q} ,t)}{\partial t} -v_{\alpha }\left (\mathbf{q} ,t\right )\right) 
\nonumber \\
& + (\delta _{\alpha \beta } -
\hat{q}_{\alpha}\hat{q}_{\beta })f_{\beta }^{u}(\mathbf{q} ,t), 
\label{deformation}
\end{align}
\begin{align}
0 = &-\eta q^{2}v_{\alpha }\left (\mathbf{q} ,t\right ) -\Gamma \left (v_{\alpha }\left (\mathbf{q} ,t\right )
-\frac{\partial u_{\alpha }(\mathbf{q} ,t)}{\partial t} \right ) 
\nonumber \\
& + (\delta _{\alpha \beta } -\hat{q}_{\alpha }\hat{q}_{\beta })f_{\beta }^{v}(\mathbf{q} ,t), 
\label{velocity}
\end{align}
where the last terms represent transverse components of the fluctuating force fields and we have introduced the 
notations $\hat{q}_{\alpha}= q_{\alpha}/ \vert \mathbf{q} \vert =q_{\alpha}/q$. 
Summation over the repeated indices will be assumed throughout this Letter.

The forces are generated by active elements, attached to the elastic network or suspended in the fluid. 
Because they are caused by internal dynamics within such elements, the net total force generated
by each element vanishes. 
However, the elements still can act as force dipoles. 
A force dipole $i$ is characterized by a unit vector $\hat{\mathbf{e}}_i$ of its orientation and by its 
magnitude $m_i$ (for a force dipole formed by a pair of forces $F_i$ and $-F_i$ at the points separated 
by distance $d_i$, we have $m_i =F_i d_i$). 
It can also be described by a nematic tensor with the elements 
$N_{i ,\alpha  \beta } =m_{i} ( \hat{e}_{i ,\alpha } \hat{e}_{i ,\beta } -\delta _{\alpha \beta }/3)$~\cite{Bruinsma14}.
Note that this nematic tensor is invariant with respect to the inversion 
$\hat{\mathbf{e}}_{i} \rightarrow  -\hat{\mathbf{e}}_{i}$ and hence the direction of the vector 
$\hat{\mathbf{e}}_{i}$ for a given line orientation can be arbitrarily assigned.  
Below we use, for convenience, the tensor $\mathbf{M}_{i}$ with the elements 
$M_{i ,\alpha \beta } =m_{i}\hat{e}_{i ,\alpha }\hat{e}_{i ,\beta }$.

We assume that the orientations $\hat{\mathbf{e}}_{i}^{u(v)}$ and positions $\mathbf{r}_{i}^{u(v)}$
of force dipoles are fixed in time, but are randomly distributed. 
Spatial distributions of the two kinds of active force dipoles are given by fluctuating tensor fields 
defined as
\begin{equation}
M_{\alpha \beta }^{u(v)}(\mathbf{r} ,t) =\sum _{i}m_{i}^{u(v)}(t)\hat{e}_{i ,\alpha }^{u(v)}\hat{e}_{i ,\beta }^{u(v)}\delta (\mathbf{r} -\mathbf{r}_{i}^{u(v)} ),
\end{equation}
which satisfies $\langle M_{\alpha\beta}^{u(v)}\rangle  =0$. 
A single active force dipole is characterized by the correlation function $S^{u(v)}(t) =\langle m^{u(v)}(t)m^{u(v)}(0)\rangle $ [see later eq.~(\ref{Suvt})].

The dipoles are statistically independent and correlation functions for their tensor fields are
\begin{align}
& \langle M_{\alpha \beta }^{u(v)}\left (\mathbf{r} ,t\right )M_{\alpha' \beta'}^{u(v)}(0 ,0)\rangle 
\nonumber \\
& = \frac{c^{u(v)}}{15}\left (\delta _{\alpha \beta}\delta _{\alpha' \beta'}+
\delta _{\alpha \alpha'}\delta _{\beta \beta'} +\delta _{\alpha \beta'}\delta _{\beta \alpha'}\right )
\delta (\mathbf{r})S^{u(v)}(t), 
\label{dipoles}
\end{align}
where $c^{u}$ and $c^{v}$ are the concentrations of force dipoles.  
We also assume that cross-correlations are absent, 
$\langle M_{\alpha \beta }^{u}M_{\alpha' \beta'}^{v} \rangle  =0$, although they can easily be included too.

The active force fields $\mathbf{f}$ can be expressed in terms of the tensor fields $\mathbf{M}$ of the dipoles. 
Assuming that each force dipole $i$ corresponds to a pair of forces $\mathbf{F}_{i}$ and 
$-\mathbf{F}_{i}$ whose application points are separated by a vector $d_{i}\hat{\mathbf{e}}_{i}$, we have
\begin{equation}
f_{\alpha }(\mathbf{r} ,t) =\sum _{i}F_{i}(t)\hat{e}_{i ,\alpha }\left[\delta (\mathbf{r} -\mathbf{r}_{i} -d_{i}\hat{\mathbf{e}}_{i}) -\delta (\mathbf{r} -\mathbf{r}_{i})\right],
\end{equation}
and therefore 
\begin{align}
&f_{\alpha }(\mathbf{q} ,t) =
\int {\rm d}\mathbf{r} \, f_{\alpha }(\mathbf{r} ,t)e^{i\mathbf{q}\cdot\mathbf{r}} \nonumber\\
=& \sum _{i}F_{i}(t)\hat{e}_{i ,\alpha }\left [\exp (i\mathbf{q}\cdot\mathbf{r}_{i} +id_{i}\mathbf{q}\cdot\hat{\mathbf{e}}_{i}) -\exp (i\mathbf{q}\cdot\mathbf{r}_{i})\right],
\end{align}
where we have dropped the superscripts $u$ or $v$.
Considering the modes with the wavelengths that are much larger than the distances between the forces, i.e.,  
$qd_{i} \ll 1$, we approximately obtain 
 \begin{equation}
 f_{\alpha }(\mathbf{q} ,t) =i\sum _{i}F_{i}(t)d_{i}q_{\beta }\hat{e}_{i ,\alpha }\hat{e}_{i ,\beta }\exp (i\mathbf{q}\cdot\mathbf{r}_{i}),
\end{equation}
and thus
\begin{equation}
f_{\alpha }^{u (v)} (\mathbf{q} ,t) =i q_{\beta } M_{\alpha  \beta }^{u (v)} (\mathbf{q} ,t) . \label{forces}
\end{equation}
Notice that, although we have derived this equation for pairs of forces, it is more general. 
On the length scales much longer than its own size, each object that asymmetrically changes its shape can 
be described as some force dipole.
Taken together, eqs.~(\ref{deformation}) and (\ref{velocity}) with the above expressions for the forces provide a complete description of the viscoelastic system with active force dipoles.

\section{Motion of tracer particles}

Reduced equations that describe stochastic motion of tracers can be derived. 
This motion is characterized by coarse-grained deformation and velocity fields that include only spatial Fourier components with relatively small wave vectors, such that $\left \vert \mathbf{q}\right \vert  <q_{\rm c}$.
The simplifying assumption is that the tracer is big, namely, that its diameter is much larger than the mesh size or, explicitly, that the condition $q_{\rm c} \ell \ll 1$ is satisfied (note that $q_{\rm c} \sim 1/a$).

Green functions $\mathbf{g}$ of the two-fluid model in eqs.~(\ref{fluid2}) and (\ref{fluid3}) have been 
derived by Levine and Lubensky~\cite{Levine00,Lubensky}.
They allow us to express deformation and velocity fields in terms of the applied forces, i.e., to write 
$u_{\alpha }(\mathbf{q} ,\omega ) =g_{\alpha \beta }^{uu}(\mathbf{q} ,\omega )f_{\beta }^{u}(\mathbf{q} ,\omega ) +g_{\alpha \beta }^{uv}(\mathbf{q} ,\omega )f_{\beta }^{v}(\mathbf{q} ,\omega )$ 
and
 $v_{\alpha }(\mathbf{q} ,\omega ) =g_{\alpha \beta }^{vu}(\mathbf{q} ,\omega )f_{\beta }^{u}(\mathbf{q} ,\omega ) +g_{\alpha \beta }^{vv}(\mathbf{q} ,\omega )f_{\beta }^{v}(\mathbf{q} ,\omega )$. 
Using these Green functions which express the forces in terms of the force dipoles through eq.~(\ref{forces}), and 
performing algebraic transformations, we obtain the following expressions for transverse components of the 
deformation and velocity fields:
\begin{align}
u_{\alpha } (\mathbf{q} ,\omega ) &  =G_{\alpha  \beta  \gamma }^{u u} M_{\beta  \gamma }^{u} (\mathbf{q} ,\omega ) +G_{\alpha  \beta  \gamma }^{u v} M_{\beta  \gamma }^{v} (\mathbf{q} ,\omega ), \\
v_{\alpha } (\mathbf{q} ,\omega ) &  =G_{\alpha  \beta  \gamma }^{v u} M_{\beta  \gamma }^{u} (\mathbf{q} ,\omega ) +G_{\alpha  \beta  \gamma }^{v v} M_{\beta  \gamma }^{v} (\mathbf{q} ,\omega ),
\end{align}
with $G_{\alpha \beta \gamma }(\mathbf{q} ,\omega ) =
ig_{\alpha \beta }(\mathbf{q} ,\omega ) q_{\gamma }$    
or, explicitly,
\begin{align}G_{\alpha  \beta  \gamma }^{u u} &  =\frac{1 +q^{2} \ell ^{2}}{\mu  q (1 +q^{2} \ell ^{2} +i \omega  \tau )} Q_{\alpha  \beta  \gamma } (\hat{\mathbf{q}})\text{,} \\
G_{\alpha  \beta  \gamma }^{u v} &  =\frac{1}{\mu  q (1 +q^{2} \ell ^{2} +i \omega  \tau )} Q_{\alpha  \beta  \gamma } (\hat{\mathbf{q}})\text{,} \\
G_{\alpha  \beta  \gamma }^{v u} &  =\frac{i \omega  \tau }{\eta  q (1 +q^{2} \ell ^{2} +i \omega  \tau )} Q_{\alpha  \beta  \gamma } (\hat{\mathbf{q}})\text{,} \\
G_{\alpha  \beta  \gamma }^{v v} &  =\frac{i \omega  \tau  +q^{2} l^{2}}{\eta  q (1 +q^{2} \ell ^{2} +i \omega  \tau )} Q_{\alpha  \beta  \gamma } (\hat{\mathbf{q}})\text{,}
\end{align}
where we have introduced $Q_{\alpha  \beta  \gamma } (\hat{\mathbf{q}}) =i \left (\delta _{\alpha  \beta } -\hat{q}_{\alpha } \hat{q}_{\beta }\right ) \hat{q}_{\gamma }$.

The above results can be simplified if only the modes with small wavenumbers $q \ell \ll 1$ are considered (but $q^2 \ell^{2}$ may still be larger than $\omega \tau $). 
By performing the inverse Fourier transform to the time-dependent deformation amplitude $\mathbf{u}(\mathbf{q} ,t)$
and the velocity amplitude $\mathbf{v}(\mathbf{q} ,t)$, reduced evolution equations for these variables can be derived.

For the deformation field, we thus find
\begin{align}
\eta \frac{\partial u_{\alpha }(\mathbf{q} ,t)}{\partial t} & = 
-\mu u_{\alpha }(\mathbf{q} ,t) 
\nonumber \\
& +\frac{Q_{\alpha \beta \gamma }(\hat{\mathbf{q}})}{q}[ M_{\beta \gamma }^{u}(\mathbf{q} ,t) +M_{\beta \gamma }^{v}(\mathbf{q} ,t)]. 
\label{deformation1}
\end{align}
On the other hand, the velocity field can be generally written as
\begin{equation}
\mathbf{v}(\mathbf{q} ,t) =\frac{\partial \mathbf{u}(\mathbf{q} ,t)}{\partial t} +\mathbf{w}(\mathbf{q} ,t), 
\label{split}
\end{equation}
separating the contribution $\mathbf{w}$ that represents the relative fluid velocity in the co-moving coordinate frame associated with the elastic component. 
For such relative velocity, the evolution equation becomes
\begin{equation}\eta \frac{\partial w_{\alpha }(\mathbf{q} ,t)}{\partial t} = 
-\mu w_{\alpha }(\mathbf{q} ,t) +\frac{\mu \ell^{2}}{\eta }qQ_{\alpha \beta \gamma }(\hat{\mathbf{q}})
M_{\beta \gamma }^{v}(\mathbf{q} ,t).
\label{velocity1}
\end{equation}
Notice here that, whereas any force dipoles present in the medium  will have an effect on the deformations in the elastic component in eq.~(\ref{deformation1}), only the force dipoles acting on the fluid component of the medium are affecting the relative velocity $\mathbf{w}$ in eq.~(\ref{velocity1}).

Equation~(\ref{deformation1}) can be used to obtain an effective equation of motion for the deformation tracer immobilized in the elastic component. The spatial position $\mathbf{U}(t)$ of such a tracer with respect to an arbitrary reference point (that can be conveniently chosen as the origin of coordinates $\mathbf{r} =0$) is given by
$U_{\alpha }(t) =(2\pi)^{-3}\int_{\left \vert \mathbf{q}\right \vert  <q_{\rm c}} {\rm d}\mathbf{q}\, 
u_{\alpha }(\mathbf{q} ,t)$.
With the use of eq.~(\ref{deformation1}), the stochastic Langevin equation for the deformation tracer is finally obtained as
\begin{equation}\eta \frac{{\rm d}U_{\alpha }}{{\rm d}t} = -\mu U_{\alpha } +F_{\alpha }^{u}(t),
\label{traceru}
\end{equation}
where the time-dependent random force $\mathbf{F}^{u}(t)$ is defined as
\begin{equation}
F_{\alpha }^{u}(t) =\int_{\left \vert \mathbf{q}\right \vert  <q_{\rm c}} 
\frac{{\rm d}\mathbf{q}}{(2\pi)^3}\, 
\frac{Q_{\alpha \beta \gamma }(\hat{\mathbf{q}})}{q}
[M_{\beta \gamma}^u(\mathbf{q},t)+M_{\beta \gamma}^v(\mathbf{q},t)].
\label{force}
\end{equation}
The correlation function of this random force is
\begin{equation}
\langle F_{\alpha }^{u}(t)F_{\beta }^{u}(0)\rangle  =\zeta^{u}q_{\rm c}[c^{u}S^{u}(t) +c^{v}S^{v}(t)]\delta _{\alpha \beta } \equiv  H^{u}(t)\delta _{\alpha \beta }, 
\label{correlationu}
\end{equation}
where the coefficient can be calculated as $\zeta ^{u}=1/(45 \pi^2)$ due to the following integral over all 
orientations of the unit vector $\hat{\mathbf{q}}$:
\begin{equation}
\int {\rm d}\hat{\mathbf{q}} \, 
Q_{\alpha \gamma \delta}(\hat{\mathbf{q}}) Q_{\beta \gamma \delta}(-\hat{\mathbf{q}})
=\frac{8\pi}{3} \delta_{\alpha \beta}.
\label{4pi}
\end{equation}

In a similar manner, a stochastic equation of motion for fluid velocity tracers can be obtained. 
According to eq.~(\ref{split}), the tracer velocity can be split into two parts, $\mathbf{V} ={\rm d}\mathbf{U}/{\rm d}t +\mathbf{W}$. 
The first part represents the velocity of a tracer if it were immobilized in the elastic component. 
This contribution to the tracer velocity can be determined by using eq.~(\ref{traceru}). 
In contrast to this, the second contribution yields the relative instantaneous velocity of the considered tracer with 
respect to the elastic component, such as the polymer network.
By using eq.~(\ref{velocity1}), we find a stochastic Langevin equation for it as 
\begin{equation}
\eta \frac{{\rm d}W_{\alpha }}{{\rm d}t} = -\mu W_{\alpha } +F_{\alpha }^{w}(t), 
\label{tracerv}
\end{equation}
where the random force is defined as
\begin{equation}
F_{\alpha }^{w}(t) =\frac{\mu \ell^{2}}{\eta }
\int_{\left \vert \mathbf{q}\right \vert  <q_{\rm c}} 
\frac{{\rm d} \mathbf{q}}{(2\pi)^3} \, q 
Q_{\alpha \beta \gamma }(\hat{\mathbf{q}})M_{\beta \gamma }^{v}(\mathbf{q} ,t).
\label{tracervv}
\end{equation}
Then its correlation function can be calculated as  
\begin{equation}
\langle F_{\alpha }^{w}(t)F_{\beta }^{w}(0) \rangle  =
\zeta ^{w}\frac{\mu ^{2}q_{\rm c}^{5}\ell^{4}}{\eta ^{2}}c^{v}S^{v}(t)\delta _{\alpha \beta },
\end{equation}
where $\zeta ^{w}=1/(225 \pi^2)$ [see also eq.~(\ref{4pi})].
This random force is determined only by the force dipoles acting on the fluid component. 
Thus, we have derived effective eqs.~(\ref{traceru}) and (\ref{tracerv}) for the motion of the two kinds 
of tracers that visualize, respectively, polymer network deformations and flows in the fluid component.

We note here that eq.~(\ref{traceru}) coincides with the phenomenological model used by Fodor 
\textit{et al.}~\cite{Fodor}. 
In their model, the spatial position $\mathbf{x}$ of a tracer particle immobilized in the network was assumed 
to follow the stochastic equation 
\begin{equation}
\gamma \frac{{\rm d}\mathbf{x}}{{\rm d}t} = -k\mathbf{x} +\mathbf{f}_{\rm A}(t), 
\label{Fodor}
\end{equation}
if we omit the thermal random forces.
In the above, $\gamma $ was the friction coefficient of the environment, $k$ was the spring constant of the surrounding network, and $\mathbf{f}_{\rm A}(t)$ was the fluctuating active force. 
To fit the model predictions to the experimental results, it was necessary to assume that $\gamma $ and $k$ were both proportional to the tracer size $a$, and that the magnitude of the active force scaled as $f_{\rm A} \sim a^{1/2}$~\cite{Fodor}.
The model described by eq.~(\ref{Fodor}) is straightforwardly obtained from eq.~(\ref{traceru}) if we identify the tracer position $\mathbf{x}$ as $\mathbf{U}$, and introduce $\gamma  =\eta /q_{\rm c} ,$ $k =\mu /q_{\rm c}$ and $\mathbf{f}_{\rm A} =\mathbf{F}^{u}/q_{\rm c}^{1/2}$, where $q_{\rm c} \sim 1/a$ as before. 
According to eq.~(\ref{force}), the active force $f_{\rm A}$ should actually have contributions coming both from the force dipoles in the elastic and fluid components of the medium.

\section{Localization vs.\ diffusion}

Using eqs.~(\ref{traceru}) and (\ref{tracerv}), one can determine time dependences of the 
mean-square displacements (MSD) for both kinds of tracers.
The analysis is simplified if we take into account that, for biological cells, the viscoelastic relaxation time 
$\tau$  can be estimated as being about 1\,ms, and this is much shorter than the typical times over which particle positions can be experimentally traced. 
Moreover, it would be typically also shorter than correlation times of active force dipoles. 
On the time scales much longer than $\tau$, the time derivative terms in eqs.~(\ref{traceru}) and (\ref{tracerv}) can be dropped and they reduce to
$U_{\alpha }(t) =F_{\alpha }^{u}(t)/\mu$ and $W_{\alpha }(t) =F_{\alpha }^{w}(t)/\mu$, respectively.
Hence, the local position of the deformation tracer and the local relative velocity of the flow tracer follow instantaneously the applied forces.
Therefore, statistical properties of such active forces should play an important role.

Various kinds of active macromolecules in a biological cell can contribute to the force dipole activity within it. 
Active force dipoles operating in the fluid subsystem are of metabolic origin.
They correspond to enzymes that undergo repeated conformational changes and thus stir mechanically the cytoplasm. 
In contrast to this, force dipoles operating in the elastic subsystem are associated with molecular motors 
such as myosin, kinesin and dynein. 
Because of their activity, the skeleton of eukaryotic cells represents an active gel. 
In bacteria and other prokaryotes, however, molecular motors are absent. 
Hence, their gel is passive and only the forces of metabolic origin are in operation in them.

Different kinds of active macromolecules in the cell will have different correlation functions of their force dipoles, and such macromolecules will be present in different concentrations.  
While we have assumed for simplicity that only one group of elastic force dipoles and one group of fluid force dipoles 
are present with their concentrations $c^{u}$ and $c^{v}$, respectively, summation over different kinds of 
macromolecules within each group has actually to be performed. 
Because of this, no universal expressions for dipole correlation functions can be formulated. 
For simple estimates, we can however assume that, within each group, dipoles are exponentially correlated in time with some correlation times $\tau ^{u}$ and $\tau ^{v}$ and we have 
\begin{equation}
S^{u(v)}(t) =\frac{S^{u(v)}}{2\tau ^{u(v)}}\exp \left ( -\frac{\vert t \vert }{\tau ^{u(v)}}\right ).
\label{Suvt}
\end{equation}
where $S^{u}$ and $S^{v}$ are integral intensities of the respective force dipoles.

The MSD for deformation tracers is obtained from eq.~(\ref{correlationu}) as
\begin{equation}
\langle \Delta U_{\alpha }^{2}(t)\rangle  =\langle \left (U_{\alpha }(t) -U_{\alpha }(0)\right )^{2}\rangle  =
\frac{2}{\mu ^{2}}[H^{u}(0) -H^{u}(t)],
\end{equation}
where $H^{u}(t)$ is the correlation function of the elastic active forces introduced 
in Eq.~(\ref{correlationu}).
In the long-time limit, or at times $t$ much longer than the correlation times $\tau ^{u}$ and $\tau ^{v}$ of all active force dipoles, we have $H^{u}(t) \rightarrow 0$. 
Hence, the MSD remains finite in this limit and localization is observed. 
The localization radius is then determined by the square root of 
\begin{equation}
\langle U_{\alpha }^{2} \rangle _{\infty } =\frac{q_{\rm c}}{15\pi^2 \mu ^{2}}
\left (\frac{c^{u}S^{u}}{\tau ^{u}} +\frac{c^{v}S^{v}}{\tau ^{v}}\right), 
\label{localization}
\end{equation}
and it decreases with the particle size as $a^{-1/2}$.

In active gels of eukaryotes, myosin motors that cross-link actin filaments are operating in large aggregates 
(microfilaments)~\cite{Misuno}. 
Each microfilament gives rise to an effective elastic force dipole and correlation times $\tau ^{u}$ for the cooperative activity of such motor groups can be long (about several seconds). 
Myosin microfilaments give rise to a time-dependent contribution into the deformation MSD at intermediate times,
$\tau ^{v} \ll t \ll \tau ^{u}$.
For such contribution, one finds that
\begin{equation}
\langle \Delta U_{\alpha }^{2}(t)\rangle  =
\frac{q_{\rm c}c^{u}S^{u}}{15 \pi^2 \mu ^{2}\tau ^{u}}\left [1 -\exp \left ( -\frac{t}{\tau ^{u}}\right )\right ] 
\approx 6 D^{u}t, 
\label{diffision-like}
\end{equation}
where 
\begin{equation}
D^{u} =\frac{q_{\rm c}c^{u}S^{u}}{90\pi^2 (\mu \tau ^{u})^{2}}.
\label{diffDu}
\end{equation}
Hence, within this time range, deformation tracers will perform an apparent diffusive motion with the ``diffusion'' coefficient $D^{u}$. 
Note that such a diffusion coefficient is inversely proportional to the particle size $a$, like the size dependence of the equilibrium diffusion coefficient given by the Stokes-Einstein relation.
The above result could previously be obtained~\cite{Fodor} in the framework of the phenomenological eq.~(\ref{Fodor}) by 
assuming  that the active force $f_{\rm A}$ satisfied the equation 
${\rm d}f_{\rm A}/{\rm d}t =\nu_{\rm A}$ where $\nu _{\rm A}(t)$ represented an active delta-correlated noise.

Next, we consider motion of flow tracers in the fluid component. Their time-dependent position vector is \begin{equation}
R_{\alpha }(t) =U_{\alpha }(t) +\int _{0}^{t} {\rm d}t_{1}\, W_{\alpha }(t_{1}),
\end{equation}
with the relative flow velocity $\mathbf{W}$ satisfying eq.~(\ref{tracerv}). 
By using this definition, the MSD for flow tracers can be obtained as
\begin{align}
\langle \Delta R_{\alpha }^{2}(t)\rangle  & =
\langle \Delta U_{\alpha }^{2}(t) \rangle  
\nonumber \\
& +\frac{1}{\mu ^{2}}\int _{0}^{t} {\rm d}t_{1}\int _{0}^{t}{\rm d}t_{2} \, 
\langle F_{\alpha }^{w}\left (t_{1} -t_{2}\right )F_{\alpha }^{w}(0)\rangle. 
\end{align}
In the long-time limit with $\tau ^{v}\ll \tau ^{u} \ll t$, we find 
\begin{equation}
\langle R_{\alpha }^{2}(t)\rangle  = \langle U_{\alpha }^{2}\rangle _{\infty } +6 D^{v}t, 
\label{diffusion1}
\end{equation}
where the diffusion coefficient is given by
\begin{equation}
D^{v} =\frac{q_{\rm c}^{5}\ell^{4} c^{v}S^{v}}{450 \pi^2 \eta ^{2}}.
\label{diffDv}
\end{equation}
Hence, the asymptotic MSD for flow tracers will include both the localization and diffusion parts. 
The diffusion is weak, but it would nonetheless dominate over the localization at sufficiently long times, 
i.e., for $t \gg \tau ^{2}/(\tau ^{v}q_{\rm c}^4 \ell^4)$.

Finally, the behavior of real physical tracers can be discussed. 
Our conjecture is that a big tracer would stochastically alternate between the two states, i.e., when it is bound to the elastic component or is free to move with the fluid component. 
Under such assumption, its observed MSD
$\langle \Delta X_{\alpha }^{2}(t)\rangle $ will be a superposition of the MSDs of deformation and fluid velocity tracers, 
\begin{equation}
\langle \Delta X_{\alpha }^{2}(t) \rangle  =\kappa \langle \Delta R_{\alpha }^{2}(t)\rangle  +
(1 -\kappa ) \langle \Delta U_{\alpha }^{2}(t) \rangle,
\label{tracers}
\end{equation}
with some weight coefficient $0 \le \kappa \le 1$.
Therefore, in the limit $t \rightarrow \infty$, we would have
\begin{equation}
\langle \Delta X_{\alpha }^{2}(t)\rangle =
\langle U_{\alpha }^{2}\rangle _{\infty} +6\kappa D^{v}t.
\end{equation}
Thus, the asymptotic motion of the tracers will represent a combination of spatial localization and of the diffusion process.

\section{Discussion}

The relationship between our theoretical results and the experimental data~\cite{Parry,Guo,Fodor} can be discussed. 
Under physiological conditions in eukaryotic cells, cooperative activity of myosin motors assembled into microfilaments acting onto the polymer network suffices to explain the main experimental results, as has been previously pointed out~\cite{Guo,Fodor}. 
The experimentally observed regime corresponds to the intermediate diffusion-like asymptotic behavior 
[see eq.~(\ref{diffision-like})] which originates from very long correlations times of the force dipoles that correspond to myosin microfilaments.

In the experiments~\cite{Guo,Fodor}, it was moreover possible to chemically inhibit myosin activity while still having ATP inside a cell. 
Additionally, tracking experiments were also performed in the same cells under depletion of ATP. 
It was found that myosin inhibition reduces the diffusion coefficient by about a factor of ten, but the remaining diffusion was still much larger than what was observed without ATP. 
Since myosin contribution has been suppressed, different mechanisms should have contributed to the remaining ATP-dependent diffusion enhancement in this case.

Note that, besides myosin, eukaryotes have other molecular motors such as kinesin and dynein in the cytoskeleton. 
However, these other motors do not form aggregates, similar to myosin microfilaments, and therefore they cannot lead to long correlation times (of the order of several seconds) that are needed to have the diffusion-like asymptotic 
behavior in eq.~(\ref{diffision-like}). 
Furthermore, strong diffusion decrease under ATP depletion (or under starvation conditions) was also observed in optical tracking experiments with bacteria that lack molecular motors in their skeleton~\cite{Parry}.
Hence the mechanisms not involving molecular motors should have been at play.

Our analysis indicates that, in viscoelastic media with active force dipoles, not only localization, but also diffusion of tracer particles will be observed in the limit of long times [see eq.~(\ref{diffusion1})].
According to eq.~(\ref{diffDv}) for $D^v$, such active diffusion is determined only by force dipoles in the fluid component. 
They should be of metabolic origin and are due to active conformational changes that accompany turnover cycles of enzymes~\cite{Parry,Kapral}. 
Note that, in accordance with eq.~(\ref{localization}), both diffusion and localization should take place.

The diffusion should prevail over the localization at the times much longer than the crossover time 
$t^{\ast} \approx \tau ^{2}/(\tau ^{v}q_{\rm c}^4 \ell^4)$. 
If the size $a \sim 1/q_{\rm c}$ of a tracer is about ten times larger than the mesh size $\ell$, and the viscoelastic 
time $\tau $ (typically about 1\,ms) is about ten times shorter than the correlation time $\tau ^{v}$ for metabolic force dipoles, we find that the crossover to classical diffusion from the saturation plateau should be after the time $t^{\ast}$ which is about 1\,s. 
This can explain why such a plateau was not seen, and only classical diffusion was found, in the experiments with bacteria where the shortest time intervals were 15\,s~\cite{Parry}.
On the other hand, a plateau of about this duration has been indeed observed in the animal cells under inhibition of myosin motors~\cite{Guo}.

It should be pointed out that obtained results hold only for relatively big tracer particles, with the sizes much larger than the mesh size (about 50\,nm) of the intracellular polymer network. 
For the tracers with the sizes smaller than the mesh size, the previous analysis in ref.~\cite{Kapral} will approximately hold.

In this study, the analysis was performed assuming idealized non-invasive flow and deformations tracers 
that serve only to visualize motions of fluid and elastic elements, without affecting flow and deformation fields themselves. 
The idealized flow tracers are therefore ``non-sticky'' and roughly correspond to those considered in a recent publication~\cite{Cai}. 
In contrast to this, deformation tracers interact strongly with the elastic component and are bound within it. 
To obtain an estimate of eq.~(\ref{tracers}), we have further assumed that the physical tracer alternates 
randomly between its free and bound states. 
It should be stressed that this is a simplification and the situation for real tracers can be much more complicated (see, e.g., ref.~\cite{Diamant15}). 
Interactions between a tracer and the elastic network depend on the ratio between the size of the tracers 
and the mesh size of the polymer network. 
When experiments with big solid tracers are performed, such tracers would certainly significantly distort 
flows of the fluid around them.

The classical two-fluid model of gels was introduced about four decades ago by de Gennes~\cite{DeGennes} 
and much progress has been made afterwards~\cite{Joanny,Fabry11}, with various detailed descriptions 
for the gels available today (see, e.g., refs.~\cite{Gustafson,Grosberg}). 
The present analysis is however done entirely within the original description~\cite{DeGennes}
in order to demonstrate what should be expected in its framework.
Remarkably, we could then reproduce the phenomenological model~\cite{Fodor} that  
shows good agreement with the experimental data for eukaryotic cells (see also the related
publication~\cite{Fodor16}). 
Extensions and generalizations of our results to other models of gels can be performed, and this 
will be the task of additional work.

The two-fluid model treats a gel as an elastic solid network with the fluid flowing through it. 
Real gels are however only partially solid and begin to flow at a finite viscosity themselves on sufficiently 
long time scales~\cite{Joanny}. 
Thus, our analysis needs to be modified at such longer time scales. 
The characteristic time when the actin gel starts to flow is of the order 100--1000\,s~\cite{Joanny}.
Note that, according to Guo \textit{et al.}~\cite{Guo}, their tweezer measurements confirmed that 
the cytoplasm is an elastic solid across the measured timescales.

Furthermore, we have not considered here the glass behavior which arises even in absence of the polymer network because the cytoplasm represents a crowded colloidal suspension. 
Previously, such behavior was investigated at thermal equilibrium~\cite{Hunter,Ooshida}.
Experimental data for bacterial cells shows that subdiffusion characteristic for glassy systems disappears in the presence of metabolic activity inside a cell and that classical diffusion is then observed~\cite{Parry}. 
Investigations of the effects of active force dipoles on crowded colloidal suspensions may be an 
interesting subject for future research.

\acknowledgments

Stimulating discussions with D.\ Mizuno are gratefully acknowledged. 
S.K.\ acknowledges support from the Grant-in-Aid for Scientific Research on
Innovative Areas ``\textit{Fluctuation and Structure}" (Grant No.\ 25103010) from the Ministry
of Education, Culture, Sports, Science, and Technology of Japan,
and the Grant-in-Aid for Scientific Research (C) (Grant No.\ 15K05250)
from the Japan Society for the Promotion of Science (JSPS).
This study was moreover supported through the EU program (Grant No.\ PIRSES-GA-2011-295243 to A.M.) 
for collaborations with Japan.


\end{document}